\documentclass[10pt,pra,noshowpacs,twocolumn,floatfix,superscriptaddress]{revtex4}  

\usepackage{graphicx}
\usepackage{epsfig}

\usepackage{amsmath,amssymb,color,float}
\usepackage{dcolumn}% Align table columns on decimal point
\usepackage{bm}% bold math

\begin{document}

\title{Thermal noise in optical cavities revisited}

\author{T. Kessler}
\address{QUEST Institute for Experimental Quantum Metrology, Physikalisch-Technische Bundesanstalt, Bundesallee 100, 38116 Braunschweig, Germany}
\email{thomas.kessler@ptb.de}
\author{T. Legero}
\address{Physikalisch-Technische Bundesanstalt, Bundesallee 100, 38116 Braunschweig, Germany}
\author{U.Sterr}
\address{Physikalisch-Technische Bundesanstalt, Bundesallee 100, 38116 Braunschweig, Germany}

%\author{T. Kessler$^{1,2,*}$, T. Legero$^2$, U. Sterr$^2$}
%\address{QUEST Institute for Experimental Quantum Metrology, Physikalisch-Technische Bundesanstalt, Bundesallee 100, 38116 Braunschweig, Germany}
%\address{Physikalisch-Technische Bundesanstalt, Bundesallee 100, 38116 Braunschweig, Germany}
%\address{Corresponding author: thomas.kessler@ptb.de}

\begin{abstract}
Thermal noise of optical reference cavities sets a fundamental limit to the  frequency instability of ultra-stable lasers. Using Levin's formulation of the fluctuation-dissipation theorem we correct the analytical estimate for the spacer contribution given by Numata \textit{et al.} \cite{num04}. For detailed analysis finite-element calculations of the thermal noise focusing on the spacer geometry, support structure and the usage of different materials have been carried out. 
We find that the increased dissipation close to the contact area between spacer and mirrors can contribute significantly to the thermal noise.   
From an estimate  of the support structure contribution we give guidelines for a low-noise mounting of the cavity. 
For mixed-material cavities we show that the thermal expansion can be compensated without deteriorating the thermal noise. 
\end{abstract}

%\ocis{%
%120.2230, % Fabry perot
% 120.3940, % Metrology
% 140.3425, % Laser stabilization
% 160.2750, % Glass and other amorphous materials
% 230.5750 % Resonators
% }

\maketitle %% null function with osajnl.sty    \maketitle

\section{Introduction}
Ultrastable optical cavities have become a standard tool for stabilizing laser systems needed e.g. for high-resolution spectroscopy, optical clocks \cite{ros08,lud08}, optical microwave generation \cite{bar05,mil09a, lip09} and coherent optical frequency transfer \cite{wil08a,jia08,ter09}. State-of-the-art cavity-stabilized laser systems show linewidths below 1 Hz and fractional frequency instabilities between $10^{-16}$ and $10^{-15}$ at one second \cite{you99,sto06,not06,lud07,web08,dub09,lod10,mil09}.
\par
The fractional length stability $\delta L/L$ of ultra-stable cavities is limited by inevitable thermal noise in the cavity materials  and a world-wide effort is ongoing to reduce this limitation.
%%%%%%%%%%%%%%%%%%%%%%% mounting krams removed
% Finally the mechanical mounting of the cavity can be optimized to minimize thermal noise. Depending on the desired application a variety of mounting schemes is used. If the cavity is intended to be used in a stationary laboratory environment it is typically mounted on a soft structure to eliminate mechanical stress \cite{web08}. 
 An efficient way of calculating thermal noise has been proposed by Levin \cite{lev98}, applying the fluctuation-dissipation theorem. Following his so-called ``direct approach'' for optical cavities, a set of  equations was given by Numata \textit{et al.} \cite{num04} for estimating  the noise contribution  of spacer, substrates and coatings.
 \par 
 The equations give a good estimate of the thermal noise arising from the mirrors which have been identified as the dominant  source of noise for cavities with mirror substrates made of the widely used ultra low expansion glass (ULE). In state-of-the-art optical cavities however, substrate materials of higher mechanical quality are being  used \cite{leg10,mil09,kes10,not95}.
For these optical cavities a fractional instability on  a level of a few $10^{-16}$ has been shown.  
At this level the spacer contribution is non-negligible and has to be calculated by numerical modeling.
In this publication we describe the mechanisms of dissipation in an optical cavity  with focus on the spacer contribution, including the support structure, the usage of different materials and the spacer geometry. We discuss current cavity designs with respect to their thermal noise budget and, where possible, will give construction guidelines for future cavity designs.
  % This publications aims for a  detailed analysis of the thermal noise created in state-of-the-art and future cavity designs.
 % Especially the estimate for the spacer contribution \cite{num04} assumes a homogeneously loaded bar which neglects  local deformations at the support points and the mirror contact surfaces.  
 % We have investigated this contribution by finite-element analysis to give more insight into this effect.
\par
In section \ref{sec:theory} we introduce the theoretical framework used for the simulations. 
We apply the simulation to a typical cavity design in section   \ref{sec:spacer} and discuss deviations from the analytical estimate for the spacer contribution. In section \ref{sec:support} we show that also the cavity support can lead to non-negligible contribution to the spacer thermal noise. In the final section \ref{sec:rings} we focus on the thermal noise floor of mixed-material cavities.
\section{Theoretical framework}
\label{sec:theory}
Thermal noise in laser mirrors has been investigated in great detail with respect to gravitational wave detection \cite{lev98,bon98,bra99,liu00,bra03,fej04,eva08,lev08a}. A variety of thermal noise sources have been identified so far \cite{gor08a}, Brownian thermal noise being the most dominant one. 
An effective way to estimate thermal noise has been first proposed  by Levin \cite{lev98}, following the fluctuation-dissipation theorem as introduced by Callen and Welton \cite{cal51}. 
The effect of Brownian motion thermal noise on the length stability of ultra-stable optical resonators has been pointed out  by Numata et al. \cite{num04} following Levin's so called ``direct approach'' \cite{lev98}. This concept is illustrated for clearness in the following.
\par
 For optical resonators the generalized coordinate describing the length fluctuations is  the averaged distance between the two cavity mirrors $x=x_1-x_2$ as probed by the laser field. According to the fluctuation-dissipation theorem, the spectral density $S_x(f)$ of the thermal fluctuations of $x$  is calculated by applying a corresponding conjugate force with amplitude $F_0$ to each mirror surface leading to  
\begin{equation}
S_x(f) = \frac{2 k_B T}{\pi^2 f^2}\frac{W_\text{diss}}{F_0^2} 
\label{eq:flucdiss}
\end{equation} 
where $T$ denotes the temperature and $k_B$ the Boltzmann constant. The term $W_\text{diss}$ denotes the time-averaged dissipated power in the system when an oscillatory  force with amplitude $F_0$ and frequency $f$ is applied. To calculate the thermal noise for a homogeneously distributed internal loss the term $W_\text{diss}$ is then expressed by 
\begin{align}
W_\text{diss}=2 \pi f U \phi
\label{eq:wdiss}
\end{align}
where $U$ denotes the maximum elastic strain energy and  and $\phi$ the loss angle of the system. 
The total dissipated power and thus the noise of the cavity length $x$ can be described as the sum of the contributions from spacer $S_x^{\text{(sp)}}(f)$, two substrates $S_x^{\text{(sb)}}(f)$ and two coatings $S^{\text{(ct)}}_x(f)$.
\begin{equation}
\begin{split}
S_x(f)&=\frac{4 k_B T}{\pi f F_0^2}\left(U_\text{sp} \phi_\text{sp}+2 U_\text{sb} \phi_\text{sb}+2 U_\text{ct} \phi_\text{ct} \right)\\
&=S_x^{\text{(sp)}}(f)+2 \cdot S_x^{\text{(sb)}}(f)+2 \cdot S_x^{\text{(ct)}}(f)
\end{split}
\label{eq:totnoise}
\end{equation}
The power spectral density of length fluctuations $S_x(f)$ can easily be converted to fractional frequency fluctuations $S_y(f)=S_x(f)/L^2$. The instability of the fluctuations in the time-domain  is conventionally characterized by the Allan-deviation $\sigma_y$ \cite{all66} of the fractional frequency fluctuations $y$. The $1/f$ noise of equation \eqref{eq:totnoise} (flicker frequency noise) leads to a constant Allan deviation
\begin{align}
 \sigma_y=\sqrt{2 \ln(2) S_y(f) f } \;.
\end{align}
In a typical  reference cavity the differential mirror displacement is probed by a Gaussian laser beam with a $1/e^2$ beam radius $w$. Consequently a  pressure distribution,
\begin{equation}
p(r)=\pm\frac{2 F_0}{\pi w^2} e^{-2 r^2/w^2}
\label{eq:pdiss}
\end{equation}
 has to be applied to the two mirrors with opposite sign to drive the excitation.  \par
 If the mirror is treated as an infinite half space the contribution of the substrate is given by \cite{lev98,bon98}
\begin{equation}
\begin{split}
U_{\text{sb}}&=\frac{1-\sigma^2}{2 \sqrt{\pi}E w} F_0^2   \\
S_x^{\text{(sb)}}(f)& =\frac{4 k_B T}{\pi f}\frac{1-\sigma^2}{2 \sqrt{\pi}E w} \phi_{\text{sb}}
\end{split}
\label{eq:usub}
\end{equation}
where $E$ denotes Young's modulus and $\sigma$ Poisson's ratio of the substrate material. 
 Bondu et al. \cite{bon98} have later refined the formula for a finite-size mirror, revealing that a long ``bar-shaped'' mirror substrate with a thickness larger than the mirror radius should be preferred to a ``gong-shaped'' substrate. However, for typical cavity geometries the beam waist is much smaller than the mirror dimensions and therefore the mirror can be approximated in good accuracy by a infinite half-space. 
\par
The coating contribution to Brownian motion thermal noise is typically estimated by treating the coating as a thin layer of thickness $d_\text{ct}$  on the substrate's front face \cite{har02b,fej04,eva08}.
Assuming a homogeneous loss angle $\phi_\text{ct}$ and similar elasticity of coating and substrate as in the case of Ti$_2$O$_5$/SiO$_2$ coatings on ULE or fused silica, the coating contribution reduces to 
\begin{equation}
\begin{split}
U_{\text{ct}}&=U_{\text{sb}} \frac{2}{\sqrt{\pi}}\frac{1-2 \sigma}{1-\sigma}\frac{d_\text{ct}}{w} \\
S_x^{\text{(ct)}}(f)&=S_x^{\text{(sb)}}(f) \frac{2}{\sqrt{\pi}}\frac{1-2 \sigma}{1-\sigma} \frac{\phi_{\text{ct}}}{\phi_{\text{sb}}}\frac{d_\text{ct}}{w}\;.
\end{split}
\label{eq:ucoat}
\end{equation}
For materials with large Young's modulus such as silicon or sapphire the thermal noise is dominated by the loss angle for strains perpendicular to the surface which is difficult to access experimentally (see discussion in \cite{har02b} for details).
\par
As a rough estimate for the spacer contribution to the cavity noise the thermal fluctuations of the length of a cylinder, averaged over the whole front face area, are calculated. A cylinder of length $L$, radius $R_{\text{sp}}$ and central bore radius $r_\text{sp}$ is assumed.
Levin's direct approach then demands a  pressure $p=F_0/A_\text{sp}$  uniformly distributed across its front faces with area $A_\text{sp}$, leading to an elastic energy of 
\begin{equation}
U_{\text{sp}}^\text{(0)}=\frac{L}{2  E A_{\text{sp}}}F_0^2=\frac{L}{2 \pi E (R_{\text{sp}}^2-r_\text{sp}^2)}F_0^2\;.
\label{eq:uspac}
\end{equation}
According to equations \eqref{eq:flucdiss} and \eqref{eq:wdiss} the thermal fluctuations of the spacer length $L$ averaged over the full cross-section of the front faces are
\begin{equation}
S_L(f)=\frac{4 k_B T}{\pi f}\frac{L}{2 \pi E (R_{\text{sp}}^2-r_\text{sp}^2)} \phi_{\text{sp}}\;.
\label{eq:sxspac}
\end{equation}
 Note, that this estimate differs from the result given by Numata~\textit{et al.} in two ways. 
Firstly, equation \eqref{eq:uspac} takes into account that the front face area of the spacer is reduced by the area of the central bore. 
Secondly, our result is a factor of $3/2$ larger than the result by Numata~\textit{et al}.  
To derive their equation the authors used the formula for the position fluctuations of one end of a free, elastic bar \cite{yam00}.  
This approach is widely employed for single mirrors in interferometers for gravitational wave astronomy.
To calculate the length fluctuations of the spacer of a Fabry-Perot interferometer the authors simply added the fluctuations from both sides which assumes
that those are uncorrelated. 
According to Levin's direct approach the calculation of the length fluctuation requires to apply opposite forces simultaneously to both ends. 
Consequently this approach automatically takes into account correlations from both ends which were neglected in \cite{num04}.
 \par

According to the model described by equation \eqref{eq:sxspac}, the noise can be minimized by distributing the stress on a large cross section. However, for mirrors smaller than the diameter of the spacer, this approximation is not valid and for a  reliable calculation of the spacer contribution of the thermal noise $S_x^\text{(sp)}$ finite element methods (FEM) have to be applied as described in the next section. 
%%%%%%%%%%%%%%%%%%%%%%%%
%%%%%%%%%%%%%%%%%%%%%%%%%%%%
\section{Simulations}
\label{sec:simulation}
\begin{figure}[t]
	\centerline{\includegraphics[width=8.4cm]{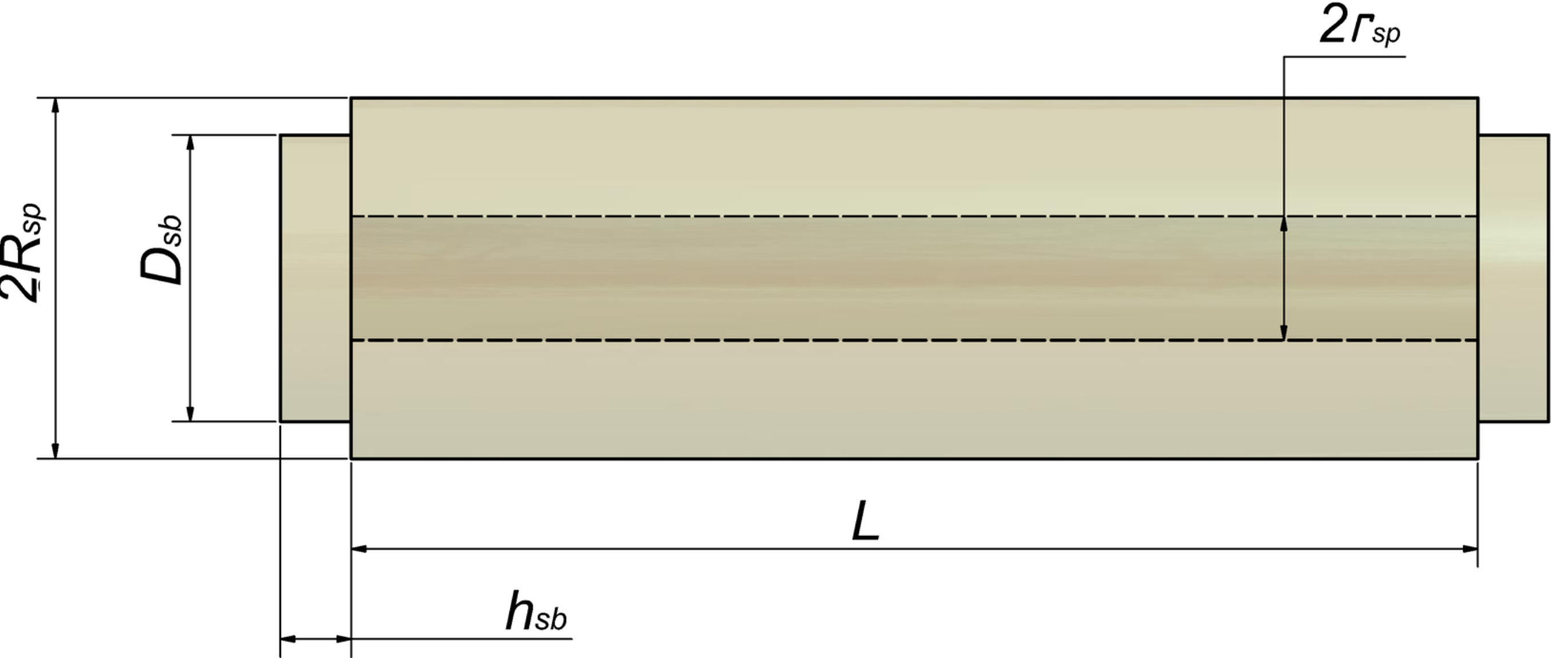}}
	\caption{Sketch of the cavity model with dimensions used for the FEM simulations in this publication}
	\label{fig:cavity}
\end{figure}

For the calculations presented in the following chapter, the software package COMSOL \cite{comsol} has been used. 
The general cavity geometry is depicted in Figure \ref{fig:cavity}.

The calculations have been carried out in axial symmetry. Because of mirror symmetry with respect to the cavity mid plane,  half a cavity was simulated. 
Figure \ref{fig:contour} shows the calculated cavity deformation under the pressure distribution of equation \eqref{eq:pdiss} assuming the parameters given in Table \ref{tab:parameter}. 
For illustration purposes a beam waist of 2 mm has been chosen.
We quote our results in terms of elastic strain energies. For ULE a strain energy of 1 nJ corresponds to a length fluctuation $S_L=8.58\cdot 10^{-35}$~m$^2/$Hz at 1 Hz and a corresponding Allan deviation of $\sigma_y=1.09\cdot 10^{-16}$ at 1 second for a cavity length of 100 mm.%%%%

The shape of the deformation of the substrate reflects the Gaussian beam profile. The majority of the elastic strain energy is stored in the cavity substrate. Strong local deformations occur at the boundary between substrate and spacer extending into the spacer. At a distance exceeding this critical depth a homogeneous energy density distribution is obtained. \subsection{Local spacer deformations}
\label{sec:spacer}
\begin{figure}[t]
\centerline{\includegraphics[width=8.4cm]{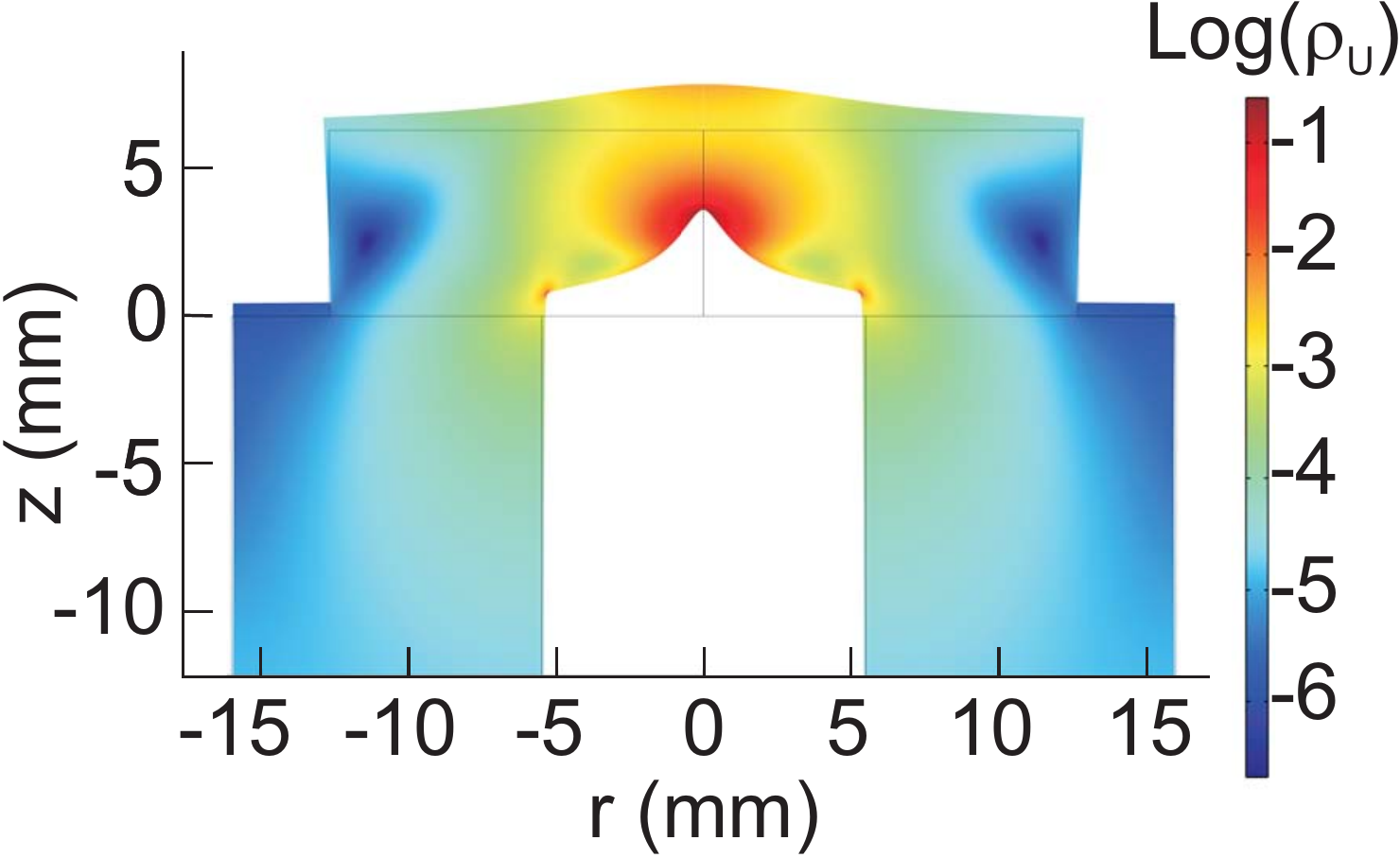}}
	\caption{Deformation and contour plot of the elastic strain energy density in the mirror substrate and spacer for a Gaussian pressure profile with a 2 mm waist on the mirror surface. The color coding corresponds to the logarithm of the energy density $\rho_\text{U}$ in SI units.}
	\label{fig:contour}
\end{figure}

Following the discussion of Figure \ref{fig:contour} it is clear that the analytic estimate for the spacer contribution according to Equation \eqref{eq:uspac} cannot hold for the full spacer length as it neglects local deformations arising from the non-uniform pressure distribution on its front face.
\par
To illustrate the size of the effect, we estimate and simulate the thermal noise following equations (\ref{eq:usub}-\ref{eq:uspac})  for an all-ULE cavity with the dimensions and material parameters given in Table \ref{tab:parameter}.

\begin{table}[b]
\small
\centerline{
\begin{tabular}{|l|r|r|}
 \hline
$R_{\text{sp}}$ & spacer radius	    &   	    16  	   mm 		\\ \hline
$L$	& spacer length  	&				   100     mm       \\ \hline
$r_\text{sp}$ &  spacer central bore radius     &   5.5  mm        \\ \hline
$D_{\text{sb}}$ & substrate diameter     &      25.4 mm               \\ \hline
$h_{\text{sb}}$ & substrate thickness     &      6.3 mm               \\ \hline
$d_{\text{ct}}$   & coating thickness						&   2   $\mu$m   \\ \hline
$w$	 &  beam waist (1/$e^2$ radius)			&		240 $\mu$m  \\ \hline\hline
$E$         & Young's modulus ULE (FS)  & 67.6 (73.0)  GPa \\ \hline
$\nu$       & Poisson ratio ULE (FS)  & 0.17 (0.16)  \\ \hline
$\phi_{\text{ct}}$ &  loss angle coating       &   $4\cdot 10^{-4}$    \\ \hline
$1/\phi$      & quality factor ULE (FS)   &    $6\cdot 10^4$ ($10^6$) \\ \hline
 $T$     &   temperature    &   293 K   \\ \hline\hline
$k_B$      & Boltzmann constant &  $1.381\cdot 10^{-23}$ J/K     \\ \hline
\end{tabular}}
\caption{Parameters used for the FEM simulation}
\label{tab:parameter}
\end{table}
%spacer: 1.5003e-009       substrate: 1.6988e-008    substrateest: 1.6885e-008         coating: 1.2702e-010      coatingest: % 1.2625e-010       spacerest: 1.0429e-009     ueberschuss: 43.8579          numata: 6.1312e-010       totalhalf: 1.7738e-008
%          dzest: 1.0429e-011

The results of the calculation are shown in Table \ref{tab:allule}.

%%%%%%
% The color coding of the contour corresponds to the energy density stored in the material on a logarithmic scale. 

\begin{table}[b]
\small
\centerline{
\begin{tabular}{|l|r|r|r|}
\hline
    & spacer   & substrate & coating \\ \hline \hline
    analytic result &  1.04   &  16.89    &   0.126      \\ \hline
    FEM calculation &  1.50   &   16.99   &   0.127 \\ \hline
\end{tabular}}
\caption{Comparison of strain energies in nJ for an all-ULE cavity}
\label{tab:allule}
\end{table}

\begin{figure}
\centering
\includegraphics[width=8.4cm]{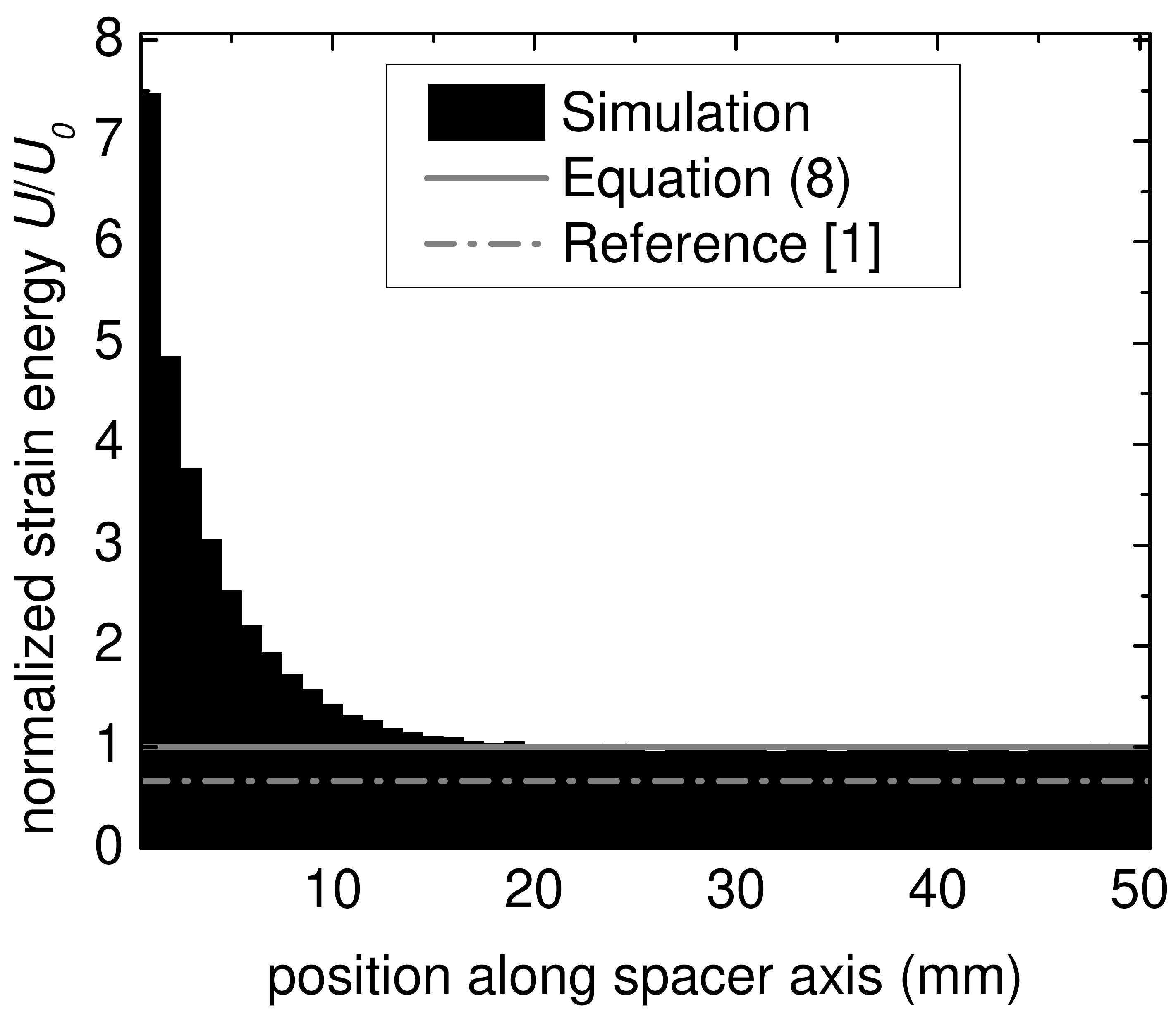}
\caption{Simulated  strain energy  in slices of 1 mm thickness compared to a homogeneously loaded spacer ($U_0$) (see Equation \eqref{eq:uspac}).  The estimate by Numata \textit{et al.}  \cite{num04} (dashed line) as well as the analytic estimate according to Equation \eqref{eq:uspac} (solid line) are shown as a reference.}
		\label{fig:hist}
\end{figure}

The major part of the total strain energy caused by the static deformation is stored in the substrates. The contribution of the substrate is in good agreement with the analytic equations  \eqref{eq:usub} on the level of 1\%. This result reflects the good approximation of the mirror surface treated as an infinite half-space, i.e. the beam waist being much smaller than the mirror dimensions. The coating contribution was  estimated by  Equation \eqref{eq:ucoat} for both the analytic calculation and the simulation.  For the spacer geometry given in table \ref{tab:parameter} Equation \eqref{eq:uspac} underestimates the energy stored in the spacer by approximately 30\%.  To shine light on this mismatch it is illustrative to calculate the energy distribution along the spacer length. The resulting graph for slices with a thickness of 1 mm is shown in Figure \ref{fig:hist}. 

The results have been normalized to the energy $U_0$ contained in a slice of a homogeneously loaded spacer, as described by equation \eqref{eq:uspac} in section \ref{sec:theory}.
While the energy stored in the central part of the spacer matches the value predicted by equation \eqref{eq:uspac}, excess energy from additional deformations is stored close to the mirror. The penetration depth amounts to roughly 10 mm. 
 The excess energy can be attributed to stress  acting non-uniformly on the front face of the spacer as illustrated in Figure \ref{fig:contour}. The analytic estimate can be reproduced by applying a homogeneous force along the spacer's front face as expected.
The amount of excess energy on the spacer ends depends on the design parameters of the cavity.
\par
\begin{figure}
\centering
				\includegraphics[width=9cm]{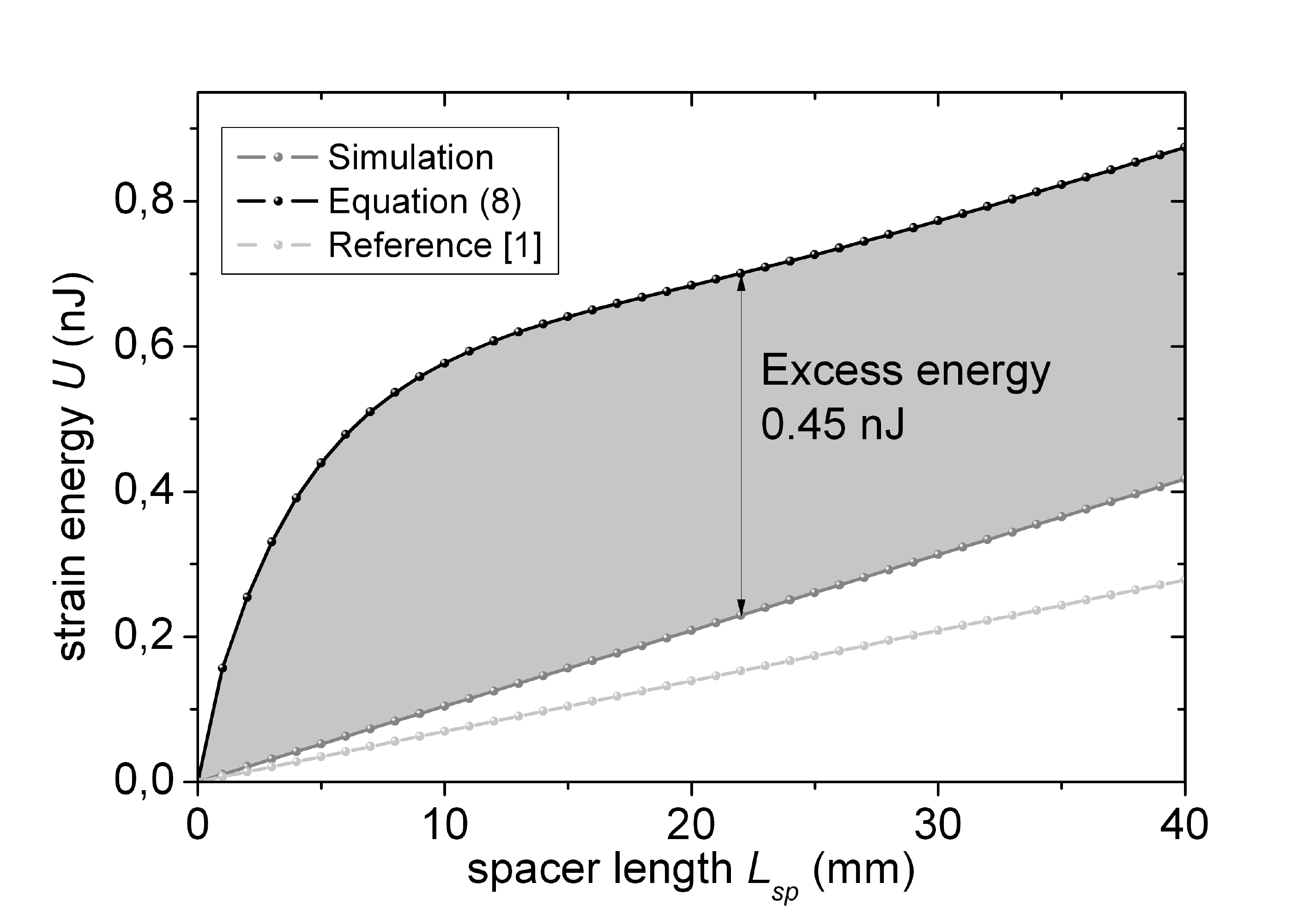}
	\caption{Strain energy as a function of spacer length for a spacer diameter of 32 mm.   The estimate by Numata \textit{et al.}  \cite{num04} (dashed line) as well as the analytic estimate according to Equation \eqref{eq:uspac} (solid line) are shown as a reference.}
		\label{fig:spacerlength}
\end{figure}

 The dependence of the strain energy on the length of the cavity is shown in Figure \ref{fig:spacerlength} in comparison with the analytic estimates from equation \eqref{eq:uspac} and from \cite{num04}.

For cavities longer than the penetration depth of the local deformation of roughly 10 mm the simulation data are offset from the analytic estimate by a constant excess energy of approximately 0.45 nJ.
Thus, once the elastic strain energy caused by the local deformations at the spacer's front faces has been determined for  a given  cavity geometry, the elastic energy of a cavity of a different length can be calculated according to equation \eqref{eq:uspac}. As the analytic estimate predicts the energy to be proportional to $L$ the effect of the local deformations becomes negligible for long spacers.
\par
\begin{figure}
\centering	
\includegraphics[width=9cm]{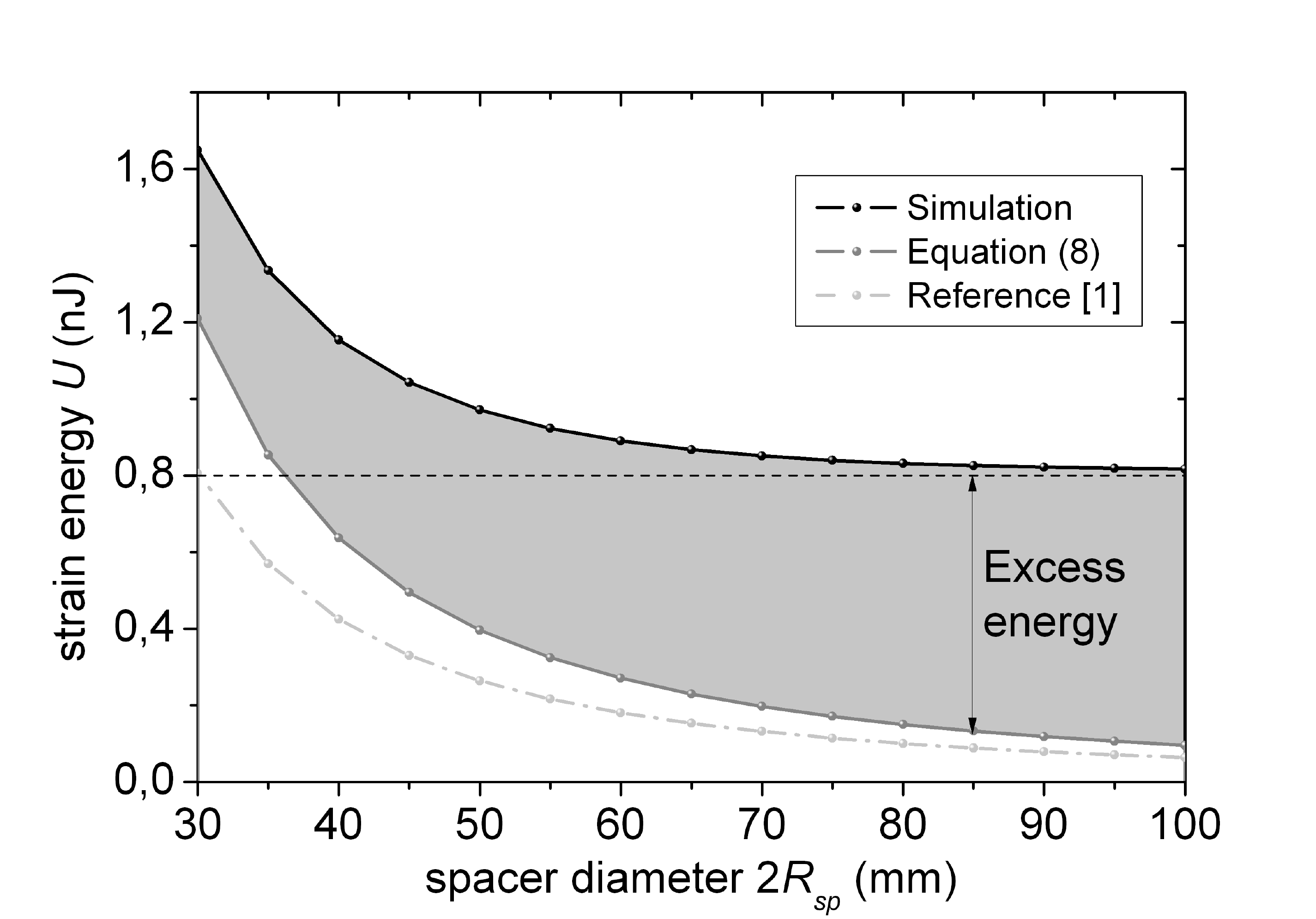}
\caption{Elastic strain energy as a function of spacer diameter for a spacer length of 100 mm.  The estimate by Numata \textit{et al.}  \cite{num04} (dashed line) as well as the analytic estimate according to Equation \eqref{eq:uspac} (solid line) are shown as a reference.}
\label{fig:spacerdiam}
\end{figure}

The dependence of the strain energy on the spacer diameter is illustrated in Figure \ref{fig:spacerdiam} 
 in comparison to the  $1/R_{\text{sp}}^2$ dependence of the analytic result (equation \eqref{eq:uspac}) and \cite{num04}. 
All three curves show a decrease of the strain energy with increasing spacer diameter. 
The excess energy increases from 0.45 nJ for a diameter of 32 mm as given in the example before and saturates at $\sim$0.8 nJ  for a spacer diameter approaching 100 mm. Note that for those diameters the spacer energy is totally dominated by the excess energy close to the mirrors ends. 
\par
In summary, equation \eqref{eq:uspac} holds for the central region of the spacer while it fails close to the spacer ends.  
The impact of the spacer contribution on the total thermal noise budget of the cavity is small for all-ULE cavities, where the dominant part of the noise arises from the mirrors.
However, it can be strongly enhanced if the spacer's mechanical losses exceed the ones for the mirrors (see section \ref{sec:rings}) or if nearly confocal cavities with large beam waists on the mirror surface lead to a reduction of the mirror thermal noise.
% beispiel??
In these cases a careful treatment of the spacer noise in a finite-element analysis is mandatory. 

\subsection{Support structure}
\label{sec:support}
\begin{figure}
\centerline{
\includegraphics[width=8.4cm]{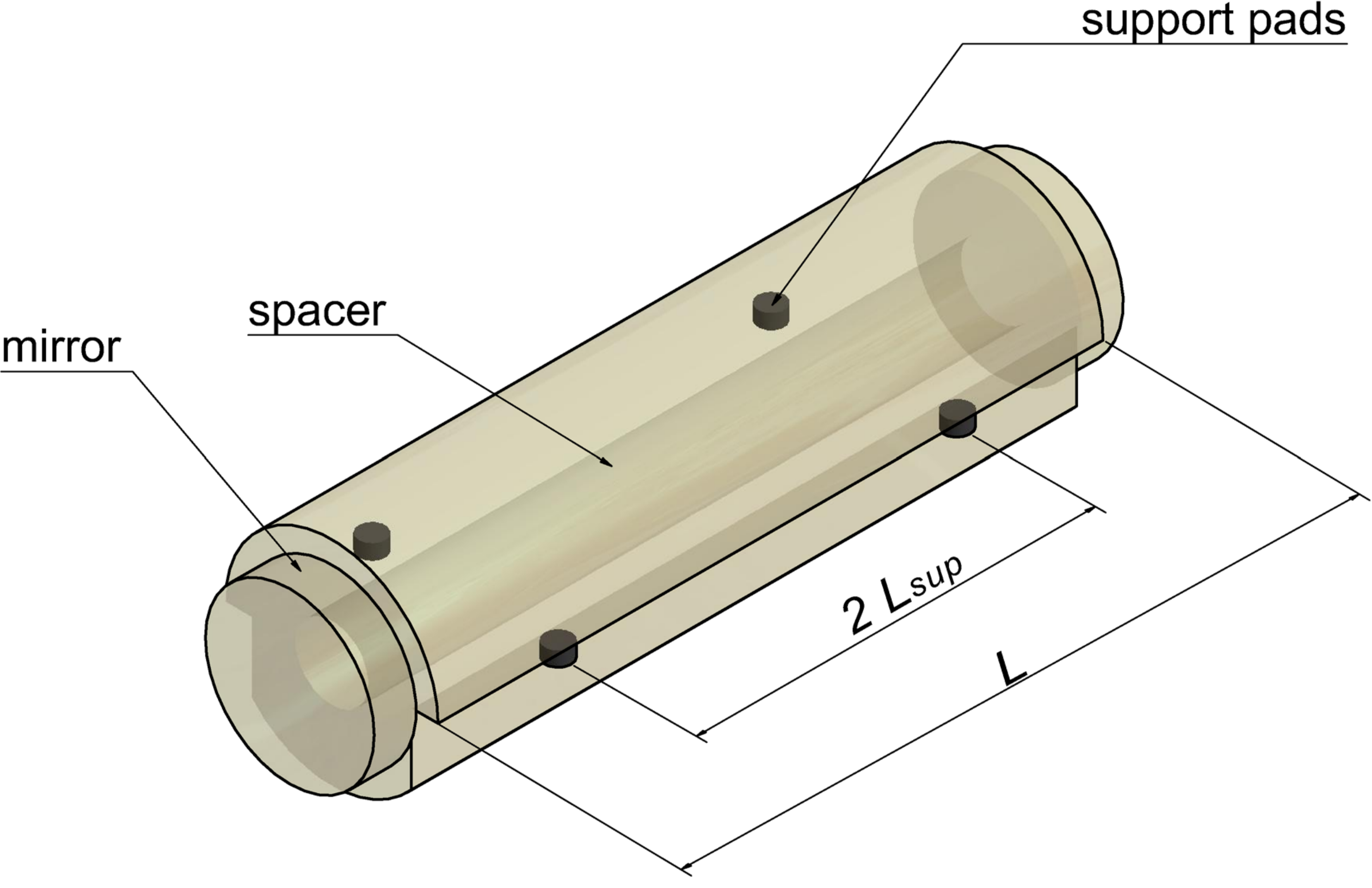} }
	\caption{Sketch of a cavity supported by four elastic support pads.}
	\label{fig:webstercavity}
\end{figure}

%%% von vorne
%%%%%%%%
To obtain fractional frequency instabilities below $\sigma_y= 10^{-15}$, modern optical cavities employ vibration insensitive mountings \cite{web10a,mil09a,web08,naz06}. In one widely used design for horizontal cavities, the resonators are  supported at four positions close to the Airy-points by elastic materials \cite{naz06,you99,web07,web08}. As the mechanical quality factor of rubber type material is relatively low the contribution of the support to the thermal noise might not be negligible. To estimate the thermal noise contribution from an elastic support we investigate the geometry depicted in Figure  \ref{fig:webstercavity}.
\par
A spacer of length $L$ is supported by four rubber supports of thickness $d_\text{sup}$, area $A_\text{sup}$ and shear modulus $G_\text{sup}$ at the Airy points positioned symmetrically at a distance $L_\text{sup}$ from the center. As the pads are located far from the spacer's front faces,  local spacer deformations as discussed in section \ref{sec:spacer} can be neglected and we can treat the spacer as a  homogeneously loaded bar. 
We again use the fluctuation-dissipation theorem in the form of equation \eqref{eq:flucdiss} to estimate the thermal noise contribution from the supports. We can estimate the energy stored in the support structure by assuming that the contact surfaces of support and spacer are displaced along the spacer axis direction according to the elastic expansion of the spacer.
The support pads are placed at a distance $L_\text{sup}$ from the center of the spacer. Opposite forces $F_0$ applied on the front faces of the spacer displace each rubber surface by 
\begin{align}
\delta L_\text{sup} =\frac{L_\text{sup}}{A_{\text{sp}} E_{\text{sp}}} F_0
\end{align} 
If we assume the support pads to be fixed to the support structure and have negligible action back on the deformation of the spacer the resulting shear energy for a support pad is given by 
\begin{equation}
\begin{split}
U_\text{sup}&=\frac{1}{2}\frac{A_\text{sup} G_\text{sup}}{d_\text{sup}} \delta L_\text{sup}^2   
   \label{eq:us}  \\ 
&\approx\frac{A_\text{sup}}{A_{\text{sp}}}\frac{L_\text{sup}}{d_\text{sup}}\frac{G_\text{sup}}{E_{\text{sp}}}\frac{L_\text{sup}}{L} U_\text{sp}^\text{(0)} \;.
\end{split}
\end{equation}
 In this approximation the shear energy is only affected by the  shear strain and the displacement of the boundary is given by the spacer displacement itself. As a result  a support with small area $A_\text{sup}$, large thickness $d_\text{sup}$ and small shear modulus $G_\text{sup}$ will reduce the strain energy and therefore the thermal noise. 
The thermal noise contribution from  four support points is then estimated by
\begin{equation}
\begin{split}
S_x^{(\text{sup})}(f) & = 4 \frac{4 k_B T}{\pi f F_0^2}U_\text{sup} \phi_\text{sup} \\
       & =4 \frac{A_\text{sup}}{A_{\text{sp}}}\frac{L_\text{sup}}{d_\text{sup}}\frac{G_\text{sup}}{E_{\text{sp}}}\frac{L_\text{sup}}{L}\frac{\phi_\text{sup}}{\phi_{\text{sp}}} S_L(f) \;.
\end{split}
\end{equation}
The distance of the support from the center can be estimated for an elastic bar  by $L_\text{sup}=L/(2 \sqrt{3})$ according to \cite{phe66}. Consequently, the displacement noise $S_{\text{sup}}(f)$ for the support is proportional to $L^2$ resulting in a fractional displacement noise $S_{y}^{\text{(sup)}}(f)=S_x^{\text{(sup)}}(f)/L^2$ independent of the cavity length. Therefore, care has to be taken especially for long spacers to avoid excessive  influence of the support on the total noise budget.
\par
For a vertical mounting of the cavity at the mid-plane  (see e.g. \cite{lud07,kes10}) the shear of the support pads acts transversal to the driving force deforming the spacer and the strain energy (equation \eqref{eq:us}) is reduced by a factor of $\left(\sigma_\text{sp}\cdot 2 R_c/L \right)^2$. 
\begin{table}
	\centerline{
		\begin{tabular}{|l|l|r|l|}
		\hline
					$A_\text{sup}$ & area & 2 mm$^2$ & \cite{mil09}\\ \hline
					$d_\text{sup}$ & thickness  & 0.7 mm & \cite{mil09}\\ \hline
					$E_\text{sup}$   & Young's modulus & 0.8 MPa &  \\ \hline
					$\nu_\text{sup}$  & Poisson's ratio & 0.27 &  \\ \hline
					$G_\text{sup}$    & shear modulus &          $E_\text{sup}/(2+2\nu_\text{sup})=0.31$ MPa & \cite{hof03}\\ \hline
					$\phi_\text{sup}$ & loss angle & 0.33 & \cite{gia96,hof03} \\ \hline
		\end{tabular}}
	\caption{Parameters for simulation of the cavity support}
	\label{tab:support}
\end{table}
\par
For the horizontal cavity design with the parameters given in tables \ref{tab:support} and \ref{tab:parameter} we can estimate that the support  
of 4 viton pads amounts to roughly 1\% of the displacement noise $S_L(f)=8.95\cdot 10^{-35}/f$ m$^2$/Hz created in an ULE spacer. For spacer materials of higher mechanical quality however  the ratio $\phi_\text{sup}/\phi_\text{sp}$ can increase drastically and therefore in future cavity designs the importance of the mounting scheme will increase. For a cavity made of silicon for example Young's modulus in the $\left[111\right]$ direction is 188 GPa \cite{wor65} and  the mechanical loss angle of the material is $\phi= 2\cdot 10^{-8}$ \cite{gui78}, about three orders of magnitude below the one for ULE (see Table \ref{tab:parameter}). Consequently, for a cavity made out of silicon the total thermal noise of the four support pads exceeds the noise of the spacer by a factor of $\sim 4$. Cavities of this type of material are currently limited entirely by the coating noise. However once low-loss coatings with sufficient reflectivities  \cite{bru10} are available this picture might change.

\subsection{Mixed material cavities}\label{sec:rings}
\begin{figure}
	\centerline{
\includegraphics[width=8.4cm]{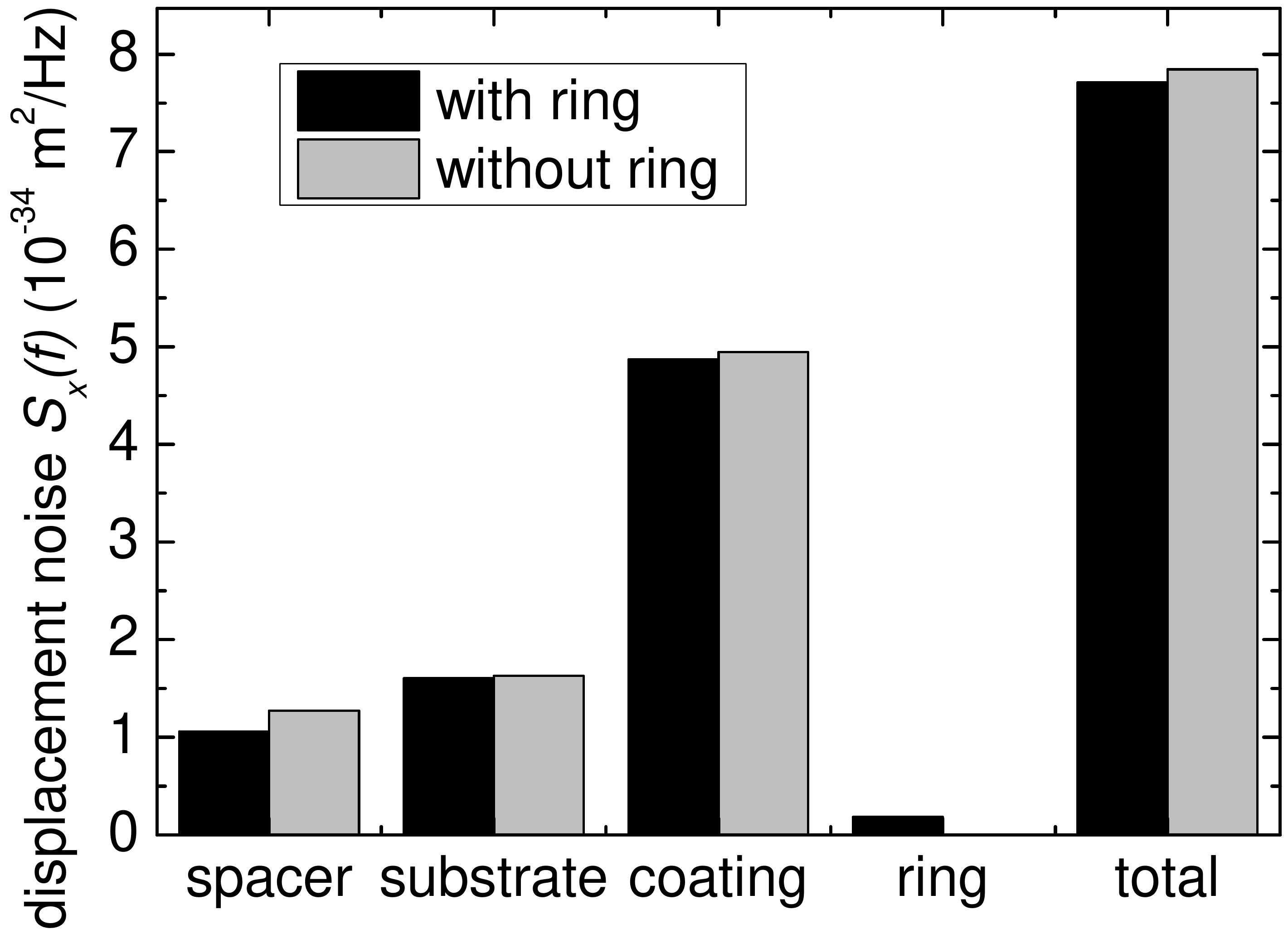}}
	\caption{Contributions to the thermal noise of the optical length  $S_x(f)$ of a cavity with and without thermal expansion compensation ring.}
	\label{fig:ring}
\end{figure}

In present cavities with ULE mirrors the major contribution to the cavity's thermal noise arises from their mirror substrates.  Substrate materials of higher mechanical Q-factor like fused silica (FS) have been used \cite{mil09,leg10} to   reduce this noise contribution. The thermal noise of FS-mirrors is already dominated by the noise of the coating (compare Figure \ref{fig:ring}). However, fused silica shows a relatively large room temperature CTE of around $5 \cdot 10^{-7}$/K. The large CTE-difference in such a mixed material cavity leads to an unwanted lowering of the zero crossing temperature of the cavity's CTE in the order of a few 10 K. To compensate for this effect, either special mirror configurations have been applied \cite{web10a} or additional ULE rings have been optically contacted to the back surfaces of the FS mirrors \cite{leg10}. 
\par
We use the FEM model discussed above to check whether the additional ULE rings of the latter approach corrupt the low thermal noise of the FS mirrors.

To the cavity model shown in Figure \ref{fig:cavity} additional  6 mm thick ULE rings with a diameter of 25.4 mm and a central bore of 9 mm have been added.  Figure \ref{fig:ring} shows the FEM results of the thermal noise contributions of the different cavity components. We find that the additional thermal noise of the ULE ring is much smaller than the noise contribution of the mirror substrate or the mirror coating. The ULE rings enlarge the effective thickness of the mirrors which results in a smaller spacer deformation and therefore in a smaller spacer excess energy. This reduction in the corresponding thermal noise is even bigger than the additional thermal noise contribution of the rings itself and we have the paradoxical situation that a combined material cavity with additional ULE rings show less thermal noise than without these rings.
\par
\begin{figure}
	\centerline{
\includegraphics[width=8.4cm]{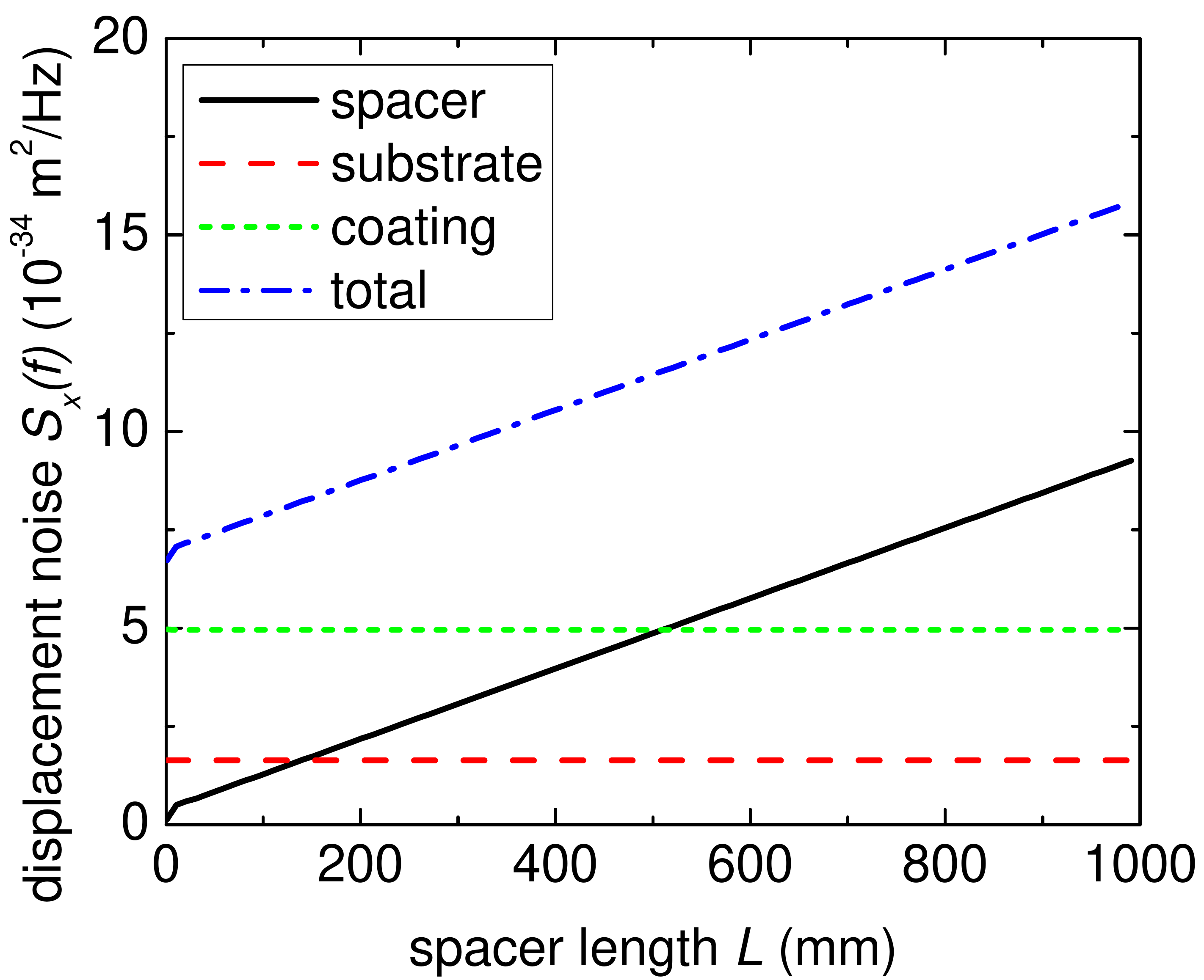}}
	\caption{Contributions to the thermal noise of the optical length $S_x(f)$ as function of the spacer length $L$.}
	\label{fig:spacerlengthfs}
\end{figure}

While for all-ULE  cavities the spacer contribution is negligible, in mixed material cavities the ULE spacer starts limiting the performance of the cavity. In general, following the discussion from section \ref{sec:theory}, the noise contribution from the spacer is proportional to $L$ and consequently a length $L$ can be found where the spacer contribution equals the mirror contribution. This effect is illustrated for a fused silica type substrate in Figure \ref{fig:spacerlengthfs}.
For a cavity with the parameters given in Table \ref{tab:parameter} this length would correspond to 500 mm.

%\begin{abstract}
%Thermal noise of optical reference cavities sets a fundamental limit to the  frequency instability of ultra-stable lasers. Using Levin's formulation of the fluctuation-dissipation theorem we correct the analytical estimate for the spacer contribution given by Numata \textit{et al} \cite{num04}. For detailed analysis finite-element calculations of the thermal noise focusing on the spacer geometry, support structure and the usage of different materials have been carried out. 
%We find that the increased dissipation close to the contact area between spacer and mirrors can contribute significantly to the thermal noise. \cite{num04}   
%From an estimate  of the support structure contribution we give guidelines for a low-noise mounting of the cavity. 
%For mixed-material cavities we show that the thermal expansion can be compensated without deteriorating of thermal noise. 
%\end{abstract}

\section{Conclusion}
\label{sec:conclusion}
Following Levin's ``direct approach'' we have analyzed the fractional length change of state-of-the-art optical cavities driven by Brownian motion thermal noise. 
While the thermal noise contribution created by the mirror is in good agreement with analytic results, we find discrepancies to the estimates given in \cite{num04} for the spacer contribution. We give a revised analytic equation valid for long spacers and find an additional contribution because of additional deformations at the spacer ends close to the mirrors.  
This excess energy can exceed the energy deposited by linear elastic deformation and can dominate the spacer contribution to the thermal noise 
for large spacer diameters and also for short spacer lengths. 
\par
Another non-negligible source for thermal noise can arise from the support structure of the material. We have given estimate equations for a cavity mounted on soft support pads from below. 
\par
Finally we have investigated the thermal noise of mixed-material cavities. We confirm that an additional compensator ring made of ULE (see \cite{leg10}) has no effect on the total cavity thermal noise. 
\par
Current cavities are mostly limited by the coating noise. However this may change as low-loss coatings are being developed like microstructured mirrors \cite{bru10} or coatings based on monocrystalline Al$_x$Ga$_{1-x}$As heterostructures \cite{col08}. Additional reduction of thermal noise may by achieved by using  large mode diameters or  higher-order transversal modes \cite{mou06}. 
In these cases the results given in this publication enable further reduction of the thermal noise from the spacer, e.g. using thick mirrors, long cavities and optimized mounting. With these precautions instabilities below $10^{-16}$ seem to be possible.
\section*{Acknowledgment}
We acknowledge financial support by the Centre for Quantum Engineering and Space-Time Research (QUEST). 
This work is supported by the European Community's ERA-NET-Plus Programme under Grant Agreement No. 217257, by the ESA and DLR in the project Space Optical Clocks.

% \bibliography{tnoise}
\bibliographystyle{osajnl}

\end{document}